\documentclass[12pt]{article}
\usepackage{latexsym,amsmath,amssymb,amsbsy,graphicx}
\usepackage[cp1251]{inputenc}
\usepackage[T2A]{fontenc}
\usepackage[russian,english]{babel}
\usepackage[space]{cite} 

\begin{document}
\bigskip

{\bf  		
GENERAL TRENDS IN THE CHANGES  OF INDICES OF SOLAR ACTIVITY IN THE LATE XX - EARLY XXI CENTURY}

\bigskip

\centerline {E.A. Bruevich $^{a}$ ,  G.V.
Yakunina $^{b}$ }

\centerline {\it $^{a,b,}$ Sternberg Astronomical Institute, Moscow
State
 University,}
\centerline {\it Universitetsky pr., 13, Moscow 119992, Russia}\

\centerline {\it e-mail:  $^a${red-field@yandex.ru},
 $^b${yakunina@sai.msu.ru}}\

\bigskip

{\bf Abstract.} The analysis of the observations of solar activity indexes SSN (NOAA Sunspot Numbers), the radio flux at a wavelength of 10.7 cm ($F_{10.7}$) and the solar constant (TSI)  during the cycles  22 - 24 is presented. We found a decrease of the observed values of the $SSN_{obs}$ which was calculated with $SSN_{syn}$ (using regression relationships between SSN and $F_{10.7}$) after 1990 year on 20 - 25\% instead of 35\%, as was previously assumed. The changes in characteristics of the most popular index, SSN, such as decrease in the number of sunspots, the reduction of the magnetic field in small and medium-sized spots are  not in full compliance with the proposed scenario of solar activity predicted by radio flux  $F_{10.7}$ in the cycles 23 and 24, and cannot be fully explained by the influence on the SSN values of additional minimum of 50 - 70 year cycle. We have also showed that the observed changes of SSN lead to a slight increase of the solar constant TSI during the cycles 23 - 24 compared to the cycle 22.

\bigskip

\bigskip

\vskip12pt
\section{Introduction}
\vskip12pt

Solar activity is concentrated mostly in active regions. It is reflected differ-ently in various layers of the solar atmosphere. Depending on the method of observation, we then observe various manifestations of this activity which, provided they have a suitable quantitative expression, we use as characteristic indices of solar activity.

Indices of solar activity 
studied in this paper are very important not only for analysis of
solar radiation which comes from  different altitudes of solar
atmosphere. 

Since all the manifestations of active regions require the presence of intensified magnetic fields, we have to assume that there is a more or less close association between their individual characteristics.

The most important for solar-terrestrial physics is the
study of solar radiation influence  on the different layers of
terrestrial atmosphere (mainly the solar constant TSI and radiation in EUV/UV
spectral range which effectively heats the thermosphere of the Earth
and so affects to our climate).

 Solar irradiance is the total amount of solar energy at a given
wavelength received at the top of the earth's atmosphere per unit
time. When integrated over all wavelengths, this quantity is called
the total solar irradiance (TSI) previously known as the solar
constant. Regular monitoring of TSI has been carried out since 1978.
From 1985 the total solar irradiance was observed by Earth Radiation
Budget Satellite (EBRS).

All the indices of solar activity are in most cases closely related, as the main source of all their variations is an variable magnetic field.
We have studied monthly averaged values of three global solar activity
indices (SSN, TSI and $F_{10.7}$) during  activity cycles 21, 22 and 23.  Most of these observed data we used in our paper were published in
Solar-Geophysical Data Bulletin (2009) and in Reports of National
Geophysical Data Center Solar and Terrestrial Physics (2014). 

Monthly averages allowed us to
take into consideration the fact that the major modulation of solar
indexes are the consequence of
 27 - 28 days variations of radiative fluxes (these variations correspond
 to the mean solar rotation period).
After monthly averaging we reduced the influence of the rotational
modulation on the observational data, see Bruevich $\it{et~ al.}$  (2014a), Bruevich and Yakunina (2015).
The various activity indices which characterized the
different aspects of the solar magnetic activity correlate quite
well with the most popular solar index such as the sunspot numbers
and with others indices over long time scales.  	

When various reconstructions of the past (and the predictions of the future) there means that the relationship between the indices of the activity remains unchanged over time. This is true for indices that are closely related, as, for example, fluxes in radio and UV ranges.

Floyd $\it{et~al.}$
(2005) showed that the mutual relation between sunspot numbers and
also between three solar UV/EUV indices, the $F_{10.7}$ flux and the
Mg II core-to-wing ratio (which is the well known chromospheric UV
index, see (Viereck $\it{et~al.}$ (2001), Viereck $\it{et~al.}$
(2004)), remained stable for 25 years until 2000. At the end of 2001
these mutual relations dramatically changed due to a large
enhancement which took place after actual sunspot maximum of the
cycle 23.
On the other hand the connection between the  radiation fluxes and indirect indices, such as the SSN  is not so obvious: processes of formation and evolution of spots are different and are poorly studied.

For all activity indices in the
$23^{rd}$ solar activity cycle one can see two maximums separated
one from another on 1,5 year approximately. We see the similar
double-peak structure in the cycle 22 but for the cycle 21 the
double-peak structure is not so evident. We see that there are
displacements in both maximum occurrence time of all these indices
in the $23^{rd}$ solar cycle. We assume that the probable reason of such double-peak structures is
a modulation of the 11-year fluxes variations by both of the
quasi-biennial and 5,5 year cyclicities of solar magnetic activity,
see Bruevich and Kononovich (2011)

Indeed, while a time there was a confidence in the close connection between the radio flux  $F_{10.7}$ and the SSN  that we can always possible to calculate the value of one from another, see Bruevich and Yakunina (2011), now we see that this relationship has being not so stable.
We have to point out that close interconnection between radiation
fluxes characterized the energy release from different atmosphere's
layers is the widespread phenomenon among the stars of late-type
spectral classes. Bruevich \& Alekseev (2007) confirmed that there
exists the close interconnection between photospheric and coronal
fluxes variations for solar-type stars of F, G, K and M spectral
classes with widely varying activity of their atmospheres. It was
shown that the sum of areas of spots and the values of X-ray fluxes
increase gradually from the Sun and HK project stars with the low
spotted discs to the highly spotted K and M-stars.

Observers believe that low values of global activity indices in 23-rd and 24-th solar cycles can be explained in recent years with help of the overlap of the current 11-year cycle with the minimum  of 50 - 70 year cycle.
The ratio of  SSN to the radio  flux $F_{10.7}$ (which is considered the most reliable solar index) from 1950 to 1990 was almost constant, and from 1990 to 2012 began to decrease parabolically 30 - 35\% (Livingston $\it{et~al.}$ (2012); Svalgaard (2013)).

The magnetic activity of the Sun is called the complex of
electromagnetic and hydrodynamic processes in the solar atmosphere
and in the under-photospheric convective zone, see Rozgacheva and
Bruevich (2002). 
Observers found that the magnetic field strength of sunspots, averaged over all spots, was gradually decreases from 1998 to 2011 by about 25\%, see Livingston $\it{et~al.}$ (2012).
If these two trends (the ratio of  SSN to the radio  flux $F_{10.7}$ decreasing simultaneously with magnetic field strength of sunspots decreasing) would develop in the same direction, the spots on the Sun would be disappeared, as in the era of the Maunder minimum, see Svalgaard (2013).

 Other indices of activity associated with surface magnetic fields also are acting by  unusual way in recent time.
 The analysis of the total area of sunspots as well as the total space of the flare regions  has made on the basis of observational data from Observatory of San Fernando (Spain), see  Chapman  $\it{et~al.}$ (2014).

  	It was shown that the relative amplitude of the total areas of the sunspots  were reduced from 1.0 in the maximum of the cycle 22 to 0.74, and 0.37  in the maximums of cycles 23 and 24 respectively.

  	Also at the Observatory of San Fernando the rations of total area of faculae  to the total area of chromospheric network (the ratio of faculae/network images in the $\it{Ca II ~K}$ line) has been measured for the cycles  22, 23 and 24.
  	In Chapman $\it{et~ al.}$ (2014) was also found that the total area of  faculae which were observed in the line of $\it{Ca II ~K}$ is reduced from the cycle 22 to the cycle 24. But in the cycle 24 the ratio of the  total areas of faculae to the total areas of spots increases.
  	
We use the TSI data set from NGDC web site http://www.ngdc.noaa.gov
and combined observational data from National Geophysical Data
Center Solar and Terrestrial Physics (2014). A significant contribution to TSI variations (Active Sun/Quiet Sun) has a UV/EUV flux that was
pointed in Krivova and Solanki (2008). Up
to 63,3 \% of TSI variability is produced at wavelengths below 400
nm. It is important that this  UV/EUV flux determines the heating of the upper thermosphere of the earth and correspondingly affects the earth's climate.
 Towards activity maxima the number of sunspots grows
dramatically. But on average the TSI is increased by about 0,1\%
from minimum to maximum of activity cycle. This is due to the
increase amounts of bright features, faculae and network elements on
the solar surface. The total area of the solar surface covered by
such features rises more strongly as the cycle progresses than the
total area of dark sunspots. Some physics-based models have been
developed with using the combined proxies describing sunspot
darkening (sunspot number or areas) and faculae brightening (faculae
areas, $\it{Ca II}$ or $\it{Mg II}$  indices), see Fontenla $\it{et~al.}$ (2004),
Krivova $\it{et~al.}$ (2003).

  	The aim of this work is the analysis of recent  general trends in the behavior of monthly averages of global indices of solar activity, such as relative spot numbers SSN, radio flux $F_{10.7}$ and the solar constant TSI in the cycles 22 - 24 .
  	
\vskip12pt
\section{Analysis of recent general trends in solar activity}
\vskip12pt  	
 
 The SSN - relative sunspot numbers index has decreased by a quarter over the past 30 years according to the observations of Penn and Livingston (2006). 

Analysis of the data of Penn and Livingston (2006), Penn and Livingston (2011) held in Nagovitsyn $\it{et ~al.}$ (2012) showed that along with the detected trend towards decreasing magnetic field spot, averaged over all spots, for the largest spots such reduction of the magnetic field is not observable but the normal variations typical of the 11-year cycle.

\begin{figure}[tbh!]
\centerline{ 
\includegraphics[width=130mm]{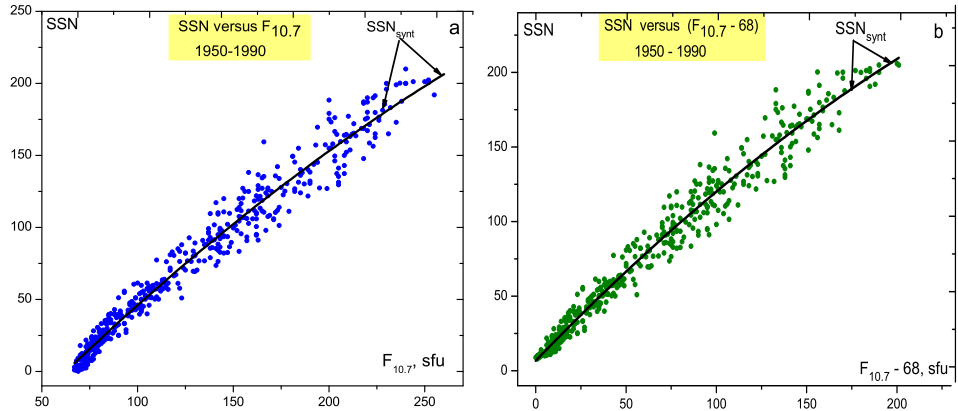}}
 \caption{Dependences:  a) - SSN from $F_{10.7}$ ; b) - SSN  from ($F_{10.7}$~-~$F_{10.7}^{~min})$.
Observations from 1950 to 1990. } 
{\label{Fi:Fig1}}
\end{figure}

In the period 1998 - 2011 the number of large spots decreased in accordance with the predictions of cyclical activity, whereas the number of small and medium-sized spots increased. 	
  	So,  a negative correlation between the number of small and big spots was found, and  an explanation for this contradiction was given.
  	It turned out that only in the separation of the spots on small and large we have the opportunity to explain the contradictions between the forecast and observations in the trends in changes of magnetic fields in spots  in the cycles 22, 23 and 24.

  	   The ratio of the SSN to the magnitude of the radiation flux on the wave of 10.7 cm remained almost constant from 1950 to 1990, after which this ratio has steadily decreased, see Livingston $\it{et~ al.}$  (2012).
  	   
For an objective assessment of this trend for observations 1950 - 1990 with  using the coefficients of the polynomial regression between SSN and the radiation flux $F_{10.7}$, the  $SSN_{syn}$ values were calculated, see Fig. 1.

  	According to our calculations of the coefficients of the polynomial regression from Fig. 1a the dependence of SSN from $F_{10.7}$ is described by the following formula:
  	
  	$$  SSN_{syn} = - 83.63 + 1.41 \cdot  F_{10.7} - 0.0011  \cdot  F_{10.7}^{~ 2}        ~ ~ ~  (1)  $$

 	Fig. 1b  shows the dependence of  SSN from 10.7 cm radio flux where, instead of a full flux
 	 $F_{10.7}$, we use only its variable part which is equal to $F_{10.7} ~- F_{10.7}^{~min}$.

It is known that the lowest value of the 10.7 cm radio flux in the minimums of 22   - 24 cycles is $F_{10.7}^{~min} = 68 ~sfu$ ($1~ sfu~= 10^{-22} ~Watt \cdot m^{-2} \cdot Hz^{-1}$), which roughly corresponds to the radiation flux of the Sun in the complete absence of spots. 

The dependence of the SSN from $(F_{10.7} - F_{10.7}^{~min})$ according to the polynomial regression from Fig. 1b is described by the following formula:

$$  SSN_{syn} = ~ 6.70 + 1.26 \cdot  (F_{10.7} - 68) - 0.0012 \cdot (F_{10.7} - 68)^2  ~ ~ ~  (2) $$

For each month of the observation based on the $F_{10.7}$ data with using the formulas (1) and (2)  the $  SSN_{syn}$ values are computed.

Then we have computed the relationship  $  SSN_{obs}$ (observed) to  $  SSN_{syn}$ (calculated). Time series of these relations are shown in Fig. 2 in accordance with the method of constructing of annual averages $  SSN_{obs}/SSN_{syn}$ according to Svalgaard (2013).

\begin{figure}[tbh!]
\centerline{ 
\includegraphics[width=130mm]{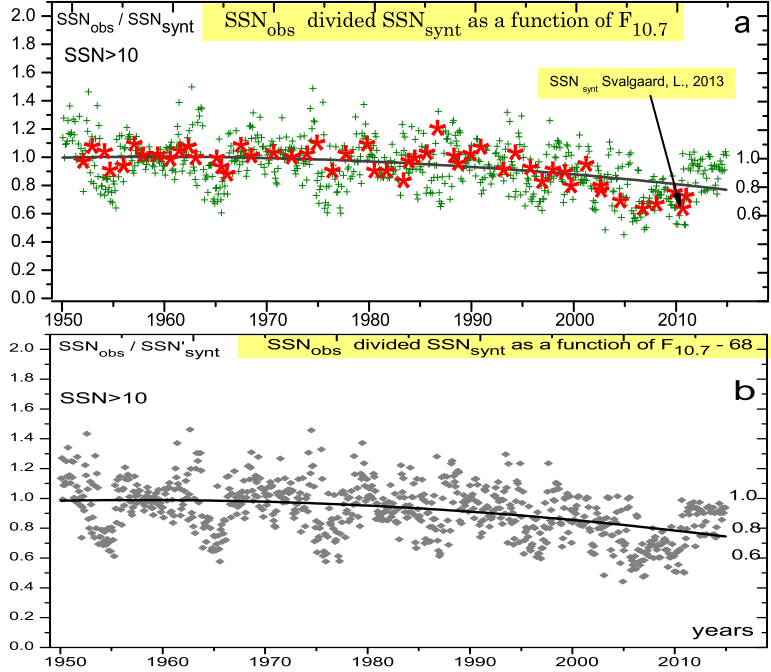}}
 \caption{a) - the time dependence of the average ratio of the observed numbers of sunspots $  SSN_{obs}$ to values $  SSN_{syn}$, calculated (crosses) using the coefficients of the polynomial regression between SSN and the radiation flux $F_{10.7}$. Annual average $SSN_{obs}$/$SSN_{syn}$ from Svalgaard (2013) are marked with asterisks; b) - the time dependence of the ratio of the observed average monthly number of sunspots $  SSN_{obs}$ to values $  SSN_{syn}$ calculated using the coefficients of the polynomial regression between SSN and the flux of radiation $(F_{10.7} - F_{10.7}^{~min})$. The months with SSN>10 are included in consideration. } 
\label{Fi:Fig2}
\end{figure}   

In Fig. 2a we show a time series of monthly averages of  $SSN_{obs}$/$SSN_{syn}$ (crosses) from 1950, extended by us until the beginning of 2015.

We can see that in 2013 - 2014, the number of sunspots increased markedly, and ultimately a reduction in the amount $SSN_{obs}$/$SSN_{syn}$ was not so great as in Livingston 
$\it{et~ al.}$ (2012); Svalgaard (2013) and is of about 20\%.

In the case where $SSN_{syn}$ is calculated depending on the variable part of the radiation flux by the formula (2) the decrease in the $SSN_{obs}$/$SSN_{syn}$ in recent years was 25\% (Fig. 2b).
Annual averages of $SSN_{obs}$/$SSN_{syn}$, shown by asterisks in Figure 2a, were taken from Svalgaard (2013).
It is seen that the Svalgaard's ratio $SSN_{obs}$/$SSN_{syn}$ has reduced by 35 \%. by 2012.
		
The difference with our results is due to the fact that in the period from 2013 to 2015, which were not included in earlier papers, the number of spots $SSN_{obs}$ has increased noticeably and the main trend was changed. Therefore the value $SSN_{obs}$/$SSN_{syn}$ decreased only by 20 - 25\% in our analysis. 

We also analysed the ratio of the  monthly averaged number of spots to the radio flux $(F_{10.7}$ taking into account only the variable components of  the radio flux: $SSN_{obs}$/$(F_{10.7} - F_{10.7}^{~min})$.

\begin{figure}[tbh!]
\centerline{ 
\includegraphics[width=130mm]{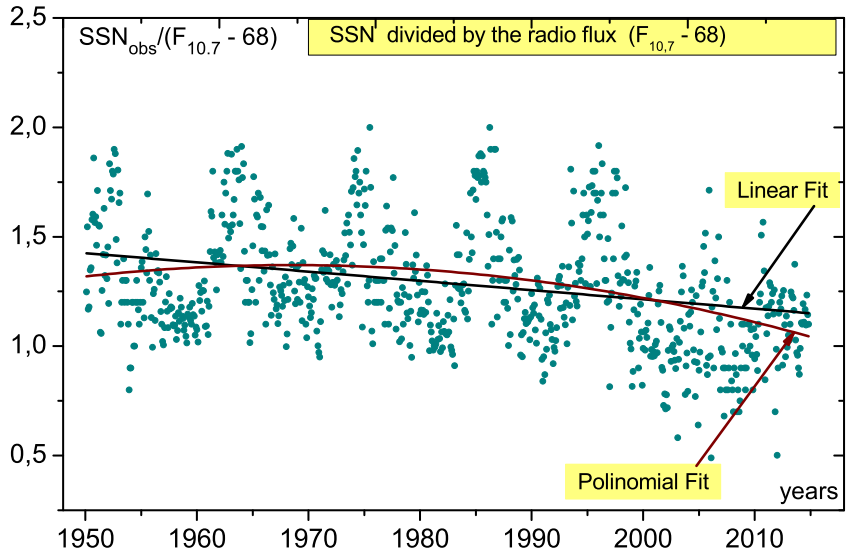}}
 \caption{The time dependence of the ratios of the observed monthly averaged values of the sunspot numbers  to the variable component of radio flux values: $SSN_{obs}$/$(F_{10.7} - F_{10.7}^{~min})$.
Observations from 1950 to 2015. } 
{\label{Fi:Fig3}}
\end{figure}

As we can  see in Fig. 3 a decrease in the relative number of sunspots after 1990, In the case of linear regression, the reduction is about 15\%, and in the case of a polynomial (it is more suitable for the analysis of cross-correlation between solar indices) - about 25\%.

Note to the cyclical nature of  $SSN_{obs}$/$(F_{10.7} - F_{10.7}^{~min})$ 
dependence.
 The deviations from regression lines reach maximum values in the minima of the 11-year cycles, when the relative error of observations of indices of activity increases. Because of this fact we have to exclude from consideration the months when 
 $SSN<10$.
 
 On the other hand, the dispersion of the deviations decreases and the relationship  $SSN_{obs}$/$(F_{10.7} - F_{10.7}^{~min})$  becomes closer to the regression line on the rise and decline phases of  the 11-year cycles, which is consistent with the findings obtained by us earlier  Bruevich 
 $\it{et~al.}$ (2014b) about the maximum correlation between solar indices during these periods. 
 
 Our analysis showed that the assumption Livingston $\it{et~ al.}$ 
(2012) and Svalgaard (2013) about the desire of the average annual values of sunspot numbers to zero, and in the near future is waiting for another Maunder minimum, is not confirmed.

We have tried to reveal the effect of increasing the radiation flux from the solar photosphere, and with it the increase in TSI as a result  of the reduction of deficit of brightness of photospheric radiation in dark spots. A comparison was made of TSI in cycle 22 with TSI in the cycle 23 and the rising phase of cycle 24  relatively to resistant solar index - radiation flux $(F_{10.7}$.

\begin{figure}[tbh!]
\centerline{ 
\includegraphics[width=130mm]{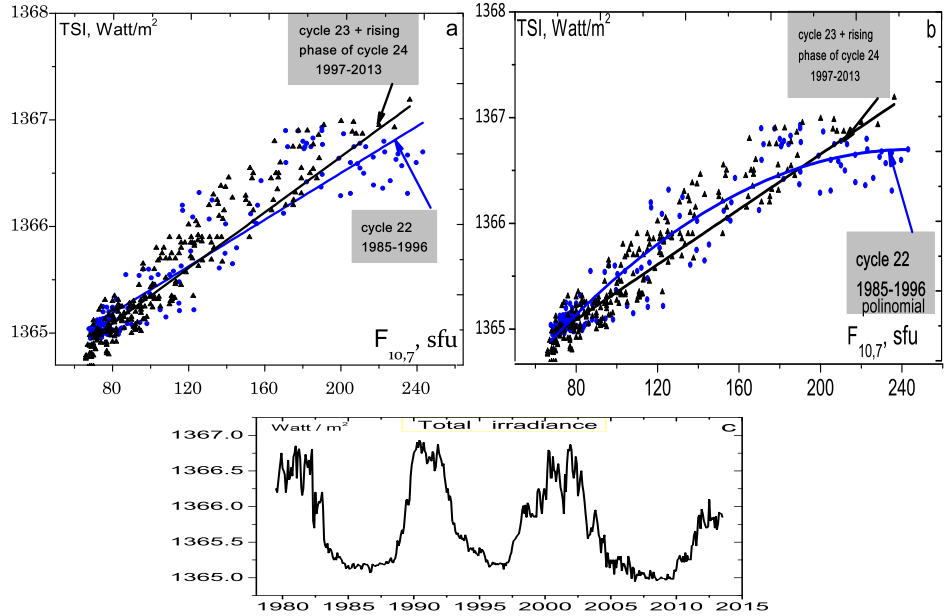}}
 \caption{The dependence of TSI from radio flux $ F_{10.7}$ for 22, and 23+24 cycles of solar activity: a) -  linear regression for 22  and (23+24) cycles; b) - linear regression for (23+24) cycles and polynomial regression for 22 cycle; c) - the set of TSI observations from 1978 to 2013. } 
{\label{Fi:Fig4}}
\end{figure}

In Fig. 4a and Fig. 4b we present the dependence of TSI from the radio  flux  at 10.7 cm. We used the observations of the total solar irradiance TSI from the entire solar disk, integrated over the entire solar spectrum. The
data of TSI monitoring from 1978 to 2013, observed by various instruments during the flight of multiple satellites that is standardized to the uniform observational  data, which are presented in National Geophysical Data Center Solar and Terrestrial Physics data of observations, see NGDC (2013).

In Fig. 4c we show  a set of observations of TSI used to analyse. Separately for the cycles 22 and 23+24 we built regression: in Fig. 4a - linear, in Fig. 4b - polynomial for the cycle 22 and linear for the cycles 23+24. It is known that polynomial regression best describes the relationship between the activity indices in the maxima of the cycles.
Fig. 4 shows that for large values of $ F_{10.7}$ the TSI values in the different cycles are different: the value of the solar constant during the periods 2000 to 2013 is higher than in 1985 - 1996.

The rms (mean-square) deviation of the TSI values from regression lines for linear and polynomial is about $0.1 ~Watt/m^2$. When $F_{10.7} = 240 ~sfu$  the difference between TSI in the 22-nd cycle and TSI in the cycles 23+24 is about $0.4 ~Watt/m^2$ for the dependencies in Fig. 4a and $0.6 ~Watt/m^2$ for the dependencies in Fig. 4b.

Thus, the difference between the two regression dependencies in Fig. 4a and in Fig. 4b are statistically significant with a significance level of 0.05 when $ F_{10.7}$ is equal to $220 ~sfu$
, and with a significance level of 0.01 in the maximum cycle when $ F_{10.7}$ is equal to
$240 ~sfu$.

This effect was predicted in Svalgaard (2013). It was assumed that the value of the solar constant should be increased when reducing  the number of sunspots and the magnetic field strength in them.

It increases the  radiation flux from the photosphere due to the decreasing of deficit of bright photosphere's flux in dark spots. Moreover, the relative amplitudes of the total areas of  spots in the cycle's maximums decrease in the  cycles 23+24 compared to the cycle 22, and the  ratio of total areas of faculae to total areas of spots in the cycle 24 is growing according to Chapman $\it{et~ al.}$ (2014). This also leads to an increase of the radiation flux from the photosphere.

It is obvious that the detection of this effect is difficult, as the maximum variation of TSI in the cycle of activity is not more than 0.1 - 0.15\% of the average  value in the cycle.

\section{Conclusions}

The analysis of relationship of mean monthly observed values $SSN_{obs}$
(NOAA Sunspots Numbers) depending on $SSN_{syn}$ from 1950 to 2015 which were calculated using the coefficients of the polynomial regression between SSN and the radio flux  $ F_{10.7}$ according to the observations 1950 - 1990, showed that for the monthly averages  $SSN_{obs}$/$SSN_{syn}$ the scatter of the deviations from the regression line is greater than in the case of yearly averages  $SSN_{obs}$/$SSN_{syn}$ from Svalgaard (2013). We also show the cyclical nature of  $SSN_{obs}$/$(F_{10.7} - F_{10.7}^{~min})$ dependence. The period of this cycle is equal to half part of 11 yr, see Fig. 3.
 The deviations from regression lines reach maximum values in the minima of the 11-year cycles, when the relative error of observations of indices of activity increases.

While an observations show marked increase in 2013 - 2014 of  the numbers of sunspots  we received a reduction in the amount  $SSN_{obs}$/$SSN_{syn}$ only 20\% (Fig. 2a), and this value is less than 35\% of Livingston $\it{et~ al.}$  (2012)and  Svalgaard (2013).

As we noted earlier in Bruevich $\it{et~ al.}$  (2014b) the relationships between the indices are much stronger in the moments related to the rising and declining phases of the 11-year cycle and worse in moments of maximums and minimums of the cycle, which is confirmed by Fig. 3.

Our detailed analysis of the variation of TSI fluxes in the cycles 22 - 24 showed that the assumption in Svalgaard (2013)  which was done on observational data of Penn and Livingston (2006), (2011) and Chapman $\it{et~ al.}$  (2014) is true: at one and the same level of  $ F_{10.7}$ the value of TSI at the cycles 23 - 24  increases, see Fig. 4.
This assumption is based on the following effect: when reducing the average number of spots and with reducing of their contrasts  the deficit of the overall flux of radiation from the solar photosphere  is slightly increased, and with it  the TSI grows.

\end{document}